\documentclass
[twocolumn,aps,prc,amssymb,amsmath,floatfix] %nofootinbib]
{revtex4-1}
\usepackage[T1]{fontenc}
\usepackage{CJK}                     % chinese fonts
\usepackage[dvips]{graphicx}         % figures
\usepackage{bm}                      % bold math
\usepackage{dcolumn}                 % align table columns on decimal point
\usepackage{mathptmx}                % math+times fonts, after bm !
\usepackage{xcolor}                  % color
\def\bea{\begin{eqnarray}} \def\eea{\end{eqnarray}}
\def\be{\begin{equation}} \def\ee{\end{equation}}
\def\bal#1\eal{\begin{align}#1\end{align}}
\def\bse#1\ese{\begin{subequations}#1\end{subequations}}
\def\rra{\right\rangle} \def\lla{\left\langle}

\def\rv{\bm{r}} \def\Jv{\bm{J}} 
\def\eps{\varepsilon}
\def\ra{\rightarrow}
\def\ccdot{\!\cdot\!}

\def\fm3{$\,\text{fm}^{-3}$}

\def\xm{\Xi^-}
\def\xim{$\Xi^-$}
\def\bxi{B_\Xi}
\def\nxi{\Xi N}
\def\ron{\rho_N} 
\def\roy{\rho_\Xi}  
\def\xbe{$^{12}_{\hskip0.27em\Xi}$Be} 
\def\xic{$^{15}_{\hskip0.27em\Xi}$C}  
\def\xib{$^{13}_{\hskip0.27em\Xi}$B}

\def\xicp{$^{15}_{\hskip0.0em\Xi p}$C}

\def\xifp{$^{21}_{\Xi p}$F}
\def\xics{$^{15}_{\hskip0.05em\Xi s}$C}
\def\xibp{$^{\hskip0.10em13}_{\Xi p}$B}

\def\xbes{$^{12}_{\hskip0.05em\Xi s}$Be}
\def\lbe{$^{10}_{\hskip0.29em\Lambda}$Be}
\def\sl{SLX}

\begin{document}

\title{Study of \xim\ hypernuclei in the Skyrme-Hartree-Fock approach}

\begin{CJK*}{UTF8}{gbsn}
%\begin{CJK*}{GB}{gbsn} % Use default fonts from CJK    emacs: raw-text
%\begin{CJK*}{GB}{SongMT}

\author{Yun Jin (金芸)}
\author{Xian-Rong Zhou (周先荣)}
\email{Corresponding author: xrzhou@phy.ecnu.edu.cn}
\author{Yi-Yuan Cheng (程弈源)}

\affiliation{
School of Physics and Materials Science, East China Normal University,
Shanghai 200241, P.~R.~China}

\author{H.-J. Schulze}
\affiliation{
INFN Sezione di Catania, Dipartimento di Fisica, Universit\'a di Catania,
Via Santa Sofia 64, 95123 Catania, Italy}

\begin{abstract}
The properties of \xim\ hypernuclei are studied systematically
using a two-dimensional Skyrme-Hartree-Fock approach
combined with three different $\nxi$ Skyrme forces
fitted to reproduce the existing data.
We explore the impurity effect of a single \xim\ hyperon
on the radii, deformations, and density distributions of the nuclear core
and point out qualitative differences between the different forces.
We find that the \xim\ removal energy of \xibp\ [$^{12}$C(g.s.)+\xim(1p)]
calculated by the \sl3 force is 0.7 MeV,
which is in good agreement with a possible value of $0.82\pm0.17\;$MeV
from the KEK E176 experiment.
The theoretical prediction for this weakly bound state
depends strongly on the deformation of the nuclear core,
which is analyzed in detail.
\end{abstract}

\maketitle
\end{CJK*}

%-------------------------------------------------------------------------------
\section{Introduction}

Hypernuclei are an important aspect of the study of nuclear structure.
In the past few decades, the investigation of hypernuclear structure
has quickly developed into a new and broad field \cite{rev}.
The motivation is to obtain clear and useful information
on the $YN$ and $YY$ interactions,
which are applied in astrophysics, for example \cite{astro}.
In 1953, the first $\Lambda$ hypernuclei were found in the laboratory
\cite{danysz}.
Apart from $\Lambda$ hypernuclei with the strangeness number $S=-1$,
there are also other hypernuclei,
such as $\Sigma$ hypernuclei ($S=-1$) and \xim\ hypernuclei ($S=-2$),
which are, however, much more difficult to produce in scattering experiments.
Therefore, only few experimental data of \xim\ hypernuclei are so far available.
The experiments for hypernuclei were conducted early at some large
experimental research facilities, such as KEK and DA$\Phi$NE.
Currently, there are several proposals for the measurement of \xim\ hypernuclei
and \xim\ atoms by new and upgraded technologies at J-PARC \cite{tanida,imai}.

Early emulsion experiments \cite{aoki,yama}
reported possible \xim\ removal energies
for the hypernucleus \xib\ [$^{12}$C+\xim] as
$3.70^{+0.18}_{-0.19}$,
$0.62^{+0.18}_{-0.19}$, or
$2.66^{+0.18}_{-0.19}$ MeV,
but it is difficult to confirm that this observation is a
bound \xim\ hypernucleus due to a lack of precise identification.
Recently, there are new events on the \xib\ hypernucleus
reported by the KEK E176 collaboration \cite{aoki2}.
A possible interpreted \xim\ removal energy is
$\bxi=0.82\pm0.17$ MeV \cite{nakazawa1,emik}.
It is expected to be consistent with a decay from the $^{12}C+\Xi^-$
system at the $2P$ state,
but this has not been confirmed.
%However, there is also the possibility that the observed event
%was a decay from an atomic $3D$ state.

Another experiment studied the production of the \xim\ hypernuclei
\xbe\ [$^{11}$B+\xim]
by the ($K^-,K^+$) reaction \cite{khaustov},
which was interpreted
by fitting a $\Xi^-$-nucleus Woods-Saxon potential with a depth of about 14~MeV.
Using this assumption,
theoretical calculations \cite{hiyama,amd} predicted values of about 4--5~MeV
for the ground-state \xim\ removal energy.

In 2015, the important ``Kiso'' event was produced
by the E373 experiment at KEK-PS \cite{nakazawa1}
in the reaction process \hbox%
{$\Xi^- + \text{$^{14}$N} \ra$ \xic\ $\ra$ \lbe\ $+\;\text{$^5_\Lambda$He}$.}
This observation is the first clear evidence of a deeply bound state of \xic.
A possible interpretation of the event
is the creation of the $p$-state hypernucleus
\xicp\ [$^{14}$N(g.s.)+\xim(1p)]
with a value of the \xim\ removal energy
$\bxi=1.11\pm0.25$ MeV \cite{nakazawa1},
later revised to
$\bxi=1.03\pm0.18$ MeV \cite{emik}.
This interpretation was shown to be consistent with the
production of \xbes\ \cite{khaustov}
within theoretical calculations in the Skyrme Hartree-Fock (SHF)
and relativistic mean field (RMF) approaches \cite{sun}.
It is also compatible with the one-peak interpretation of the
preliminary analysis
of \xbes\ production in the J-PARC E05 experiment \cite{e05}. %6.3MeV
The strongly attractive nature of the $\nxi$ interaction (XNI)
has also recently been confirmed by the
latest experimental analysis on proton-\xim\ correlations
performed by the ALICE Collaboration \cite{sacharya}.

Due to insufficient experimental information on the XNI,
it is important to study \xim\ hypernuclei
in different theoretical approaches \cite{emik}:
Within the framework of the cluster model,
the light hypernucleus \xbe\ was calculated based on the assumption
of an effective \xim\ potential whose depth is about 14 MeV.
It was predicted that the $\Xi^-$ removal energy of the \xbe\ ground state
is about 5 MeV and 2.2 MeV, with/out the $\nxi$ Coulomb interaction,
respectively \cite{hiyama}.
The RMF method performed studies on spherical \xim\ hypernuclei, such as
$^{17}_{\ \Xi}$N, $^{41}_{\ \Xi}$K, $^{91}_{\ \Xi}$Y, $^{209}_{\ \ \ \Xi}$Tl,
etc.~\cite{mares,tan,liu},
where it was found
that the \xim\ binding energies increase with the mass number of hypernuclei
and that the $\nxi$ Coulomb interaction makes the hyperon potential much deeper.
Within the quark-mean-field approach,
systematic studies were performed on \xim\ hypernuclei from
$^{12}_{\ \Xi}$Be to $^{209}_{\ \ \ \Xi}$Tl \cite{hu}.
The XNI parameters were determined under the assumption
of a \xim\ potential depth at saturation density
$U_\Xi=-12$ or $-9$ MeV.
In the framework of a one-dimensional (1D) SHF model,
three different XNIs \sl0, \sl2, \sl3 were proposed in \cite{sun}
to reproduce in a consistent manner the experimental data for \xbe\ and \xic\
discussed above.

In this work we continue and extend the study of \xim\ hypernuclei
in a two-dimensional (2D) SHF model.
One aim is to explore the impurity effect of a single \xim\ hyperon
on the deformation of core nuclei.
We perform systematic calculations of \xim\ hypernuclei from
$^{9}_{\Xi}$Li to $^{209}_{\ \ \ \Xi}$Tl,
in particular for a series of hypernuclei with deformed cores
from $^{9}_{\Xi}$Li to $^{37}_{\ \Xi}$Cl.
We focus on the \xim\ binding energies, deformations, and
nucleon and hyperon density distributions of \xim\ hypernuclei,
in particular study the dependence of the theoretical predictions
on the different XNIs \sl0, \sl2, \sl3.

The paper is organized as follows.
In Sec.~\ref{s:int} we present the method of the self-consistent 2D SHF model
and introduce the Skyrme forces for the effective XNI.
Sec.~\ref{s:res} shows the results and discussion of
radii, deformations, binding energies,
and nucleon and hyperon density distributions
for \xim\ hypernuclei from $^{9}_{\Xi}$Li to $^{209}_{\ \ \ \Xi}$Tl.
In Sec.~\ref{s:end}, we summarize the paper.

%-------------------------------------------------------------------------------
\section{Formalism}
\label{s:int}

Our approach is the 2D SHF model,
which is combined with a density-dependent Skyrme force for the XNI.
In the self-consistent model,
the total energy of a hypernucleus is calculated as
\cite{vaut,vautdef,rayet,shfrev1,hypsky}
\be
 E = \int d^3\rv\, \eps(\rv) \ ,\quad
 \eps = \eps_{NN} + \eps_{\nxi} + \eps_C \:,
\label{e:shf}
\ee
where $\eps_{NN}$ is the energy density of the nucleon-nucleon interaction,
$\eps_{\nxi}$ is the contribution due to the $\nxi$ interaction, and
$\eps_C$ is the energy density of the Coulomb interaction between protons and
\xim\ hyperon.
These energy-density functionals
are dependent on the one-body density $\rho_q$,
kinetic density $\tau_q$, and spin-orbit current $\Jv_q$,
\be
  \big[ \rho_q,\; \tau_q,\; \Jv_q \big] =
  \sum_{i=1}^{N_q} {n_q^i} \Big[
  |\phi_q^i|^2 ,\;
  |\bm\nabla\phi_q^i|^2 ,\;
  {\phi_q^i}^* (\bm\nabla \phi_q^i \times \bm\sigma)/i
 \Big] \:,
\ee
where $\phi^i_q$ ($i=1,N_q$) are the
self-consistently calculated single-particle (s.p.) wave functions
of the $N_q$ occupied states for the different particles
$q=n,p,\Xi$ in a hypernucleus.
The minimization of the total energy in Eq.~(\ref{e:shf})
implies the SHF Schr\"odinger equation for each s.p.~state $(i,q)$,
\be
 \Big[ -\bm\nabla\cdot\frac{1}{2m^*_q(\rv)}\bm\nabla
 + V_q(\rv) - i{\bm W}_q(\rv)\cdot(\bm\nabla\times\bm\sigma)
 \Big] \phi_q^i(\rv)
 = e_q^i\phi_q^i(\rv) \:,
\label{two}
\ee
where the mean fields of nucleons and hyperon
(including the Coulomb interaction)
are written as
\bea
 V_q &=& V_q^\text{SHF} + V_q^{(\Xi)} \ ,\
 V_q^{(\Xi)} = \frac{\partial\eps_{\nxi}}{\partial\rho_q}
 \ ,\ (q=n,p) \:,
\\
 V_\Xi &=& \frac{\partial\eps_{\nxi}}{\partial\roy} - V_C \:.
\eea
The nucleonic spin-orbit mean field is represented by $\bm W_{n,p}$
and is provided by the $NN$ Skyrme force used here,
whereas we assume $\bm W_\Xi=0$ in this work.
For the nucleonic energy density functional $\eps_{NN}$
we employ the SLy4 parametrization \cite{shfrev1,sly},
while the energy density functional for the hyperonic part is given by
\cite{rayet,hypsky,sun},
\bea
 \eps_{\nxi} &=& \frac{\tau_\Xi}{2m_\Xi}
  + a_0 \roy\ron
  + a_3 \roy\ron^2
  + a_1 (\roy\tau_N + \ron\tau_\Xi)
\\\nonumber
 && -a_2 (\roy\Delta\ron+\ron\Delta\roy)/2
    -a_4 (\roy\bm\nabla\ccdot\Jv_N + \ron\bm\nabla\ccdot\Jv_\Xi) \:,
\eea
which provides the hyperonic SHF mean fields
\bea
 V_{\Xi} &=& a_0\ron + a_3\rho^2_N
 + a_1\tau_N - a_2\Delta\ron - a_4\bm\nabla\ccdot\Jv_N
\:,
\label{e:vy}
\\
 V^{(\Xi)}_N &=&
 a_0\roy  + 2a_3\ron\roy
 + a_1\tau_\Xi - a_2\Delta\roy - a_4\bm\nabla\ccdot\Jv_\Xi
\:,
\label{e:vnx}
\eea
and a $\Xi$ effective mass
\be
 \frac{1}{2m^*_\Xi} = \frac{1}{2m_\Xi} + a_1\ron \:.
\ee
The relation to the standard $\nxi$ Skyrme parameters $t^{\nxi}_{0,1,2,3}$ is
\be
 a_0=t_0\ ,\
 a_1=\frac{t_1+t_2}{4}\ ,\
 a_2=\frac{3t_1-t_2}{8}\ ,\
 a_3=\frac{3t_3}{8} \:.
\ee

Due to lack of experimental data for \xim\ hypernuclei,
three simple $\nxi$ Skyrme forces \sl0, \sl2, \sl3
that employ no more than two parameters were proposed in Ref.~\cite{sun}.
(We use here the notation SLX* $\equiv$ SL*$_p$ of that work).
The \sl0 force involves only a volume term $\sim a_0$,
while in \sl2 and \sl3 the parameters are chosen as
$a_2=20\;\text{MeVfm}^5$ and
$a_3=1000\;\text{MeVfm}^6$, respectively,
motivated by $N\Lambda$ Skyrme forces \cite{hypsky}.
The values of $a_0$ in \sl0, \sl2, \sl3
were fixed to $-128,-138,-228\;\text{MeVfm}^3$, respectively,
reproducing the experimental data of $\bxi=1.11\pm0.25$ MeV
in the hypernucleus \xicp\
with the \xim\ hyperon occupying the $1p$ orbit.
(We do not refit the slightly revised value $\bxi=1.03\pm0.18$ MeV \cite{emik}
in order to be consistent with our previous work \cite{sun}).
The \sl3 force then predicts also results
compatible with the experimental information on \xbes\ \cite{khaustov}.

In our approach, we assume axial symmetry of the mean field,
and the deformed SHF Schr\"odinger equation is solved in
cylindrical coordinates $(r,z)$ within the
axially-deformed harmonic-oscillator basis \cite{vautdef,shfrev1}.
This allows us to make calculations of 2D-deformed nuclei and hypernuclei.
The quadrupole deformation parameter of the nuclear core is expressed as
\be
 \beta_2 \equiv \sqrt{\pi\over5}
 {\langle 2z^2-r^2 \rangle \over \langle r^2 + z^2 \rangle} \:,
\label{e:beta2}
\ee
while the nuclear core radius is given by
\be
 R_N \equiv \sqrt{ \lla r^2 + z^2 \rra }
 = \sqrt{ \frac{N}{A-1} \lla R_n^2 \rra + \frac{Z}{A-1} \lla R_p^2 \rra} \:.
\label{e:rn}
\ee
The calculated results of these observables will be discussed
in the next section,
together with the $\xm$ separation energy
\be
% \bxi \equiv E[^AZ] - E[^{A+1}_{\quad\Xi}(Z-1)] \:,
 \bxi \equiv E[^{A-1}(Z+1)] - E[^A_{\Xi}Z] \:,
\ee
in which notation $Z$ is the charge number
(not the proton number)
and $A$ is the total baryon number of the hypernucleus.

At the present stage of investigation
we neglect the difference of the $\Xi^-p$ and $\Xi^-n$ interactions
for the nearly symmetric nuclei we are considering here.
Also, the imaginary part of the \xim\ potential in our approach
is currently disregarded,
because it is fairly small at low momenta,
$\text{Im}\,U(k=0)/\text{Re}\,U(k=0) \approx 0.2$
in Brueckner-Hartree-Fock (BHF) calculations \cite{hypesc},
where $\text{Im}\,U$ has a strong dependence on the coupling of the
$\nxi$ and $\Lambda\Lambda$ channels.

\begin{table*}[t]%...............................................................
\caption{
Deformation parameter $\beta_2$,
separation energy $\bxi$,
and nuclear core radius $R_N$ for
several hypernuclei obtained with the \sl0, \sl2, and \sl3 forces.
The subscripts $s$ and $p$ denote the orbit that the $\xm$ hyperon occupies.
The $\bxi$ values in brackets are the results of deformed SHF calculations
without $\nxi$ Coulomb interaction.
Experimental data \cite{aoki2,nakazawa1},
and theoretical results of a cluster model calculation \cite{hiyama}
and a RMF model \cite{hu}
are listed for comparison.
}
%\mycom{are 0.001, 0.003, 0.004 numerical artefacts ?}
\def\myc#1{\multicolumn{1}{c|}{$#1$}}
\def\myt#1{\multicolumn{1}{c}{#1}}
\setlength{\tabcolsep}{0.1pt}
\renewcommand{\arraystretch}{1.0}
\vspace{1mm}
\begin{ruledtabular}
\begin{tabular}{l|r|rrrr|rr|rr|rr|c|cccc}
&\myc{K^\pi} & \multicolumn{4}{c|}{$\beta_2$}
&\multicolumn{7}{c|}{$\bxi$ (MeV)}
&\multicolumn{4}{c}{$R_N$ (fm)} \\
%\hline
& & \myt{no $\Xi$} & \myt{\sl0} & \myt{\sl2} & \multicolumn{1}{c|}{\sl3}
& \multicolumn{2}{c|}{\sl0} & \multicolumn{2}{c|}{\sl2} & \multicolumn{2}{c|}{\sl3}
& Data & \myt{no $\Xi$} & \myt{\sl0} & \myt{\sl2} & \myt{\sl3} \\
\hline

$^{\ 9}_{\Xi s}$Li & $0^+\otimes{\Xi\frac{1}{2}}^+$
& 0.623 & 0.590 & 0.592 & 0.616 & 3.8 & (1.7) & 3.2 & (1.1) & 3.6 & (1.0)
& & 2.548 & 2.477 & 2.492 & 2.558 \\

$^{10}_{\Xi s}$Li&${n\frac{3}{2}}^-\otimes{\Xi\frac{1}{2}}^+$
& 0.309 & 0.244 & 0.250 & 0.320 & 5.3 & (3.0) & 4.6 & (2.4) & 4.1 & (2.0)
& 3.6 (1.6) \cite{hiyama} & 2.438 & 2.368 & 2.382 & 2.460 \\

$^{12}_{\Xi s}$Be&${p\frac{3}{2}}^-\otimes{\Xi\frac{1}{2}}^+$
& 0.125 & 0.100 & 0.109 & 0.128 & 8.0 & (5.1) & 7.2 & (4.4) & 5.2 & (2.6)
& 5 (2.2) \cite{hiyama} & 2.439 & 2.387 & 2.397 & 2.459 \\

$^{13}_{\Xi s}$B&${p\frac{3}{2}}^-\otimes{\Xi\frac{1}{2}}^+$
&-0.302 &-0.254 &-0.254 &-0.295 & 8.1 & (4.8) & 7.6 & (4.3) & 6.2 & (3.2)
& & 2.598 & 2.523 & 2.530 & 2.606 \\

$^{13}_{\Xi p}$B&${p\frac{3}{2}}^-\otimes{\Xi\frac{3}{2}}^-$
&-0.302 &-0.311 &-0.310 &-0.318 & 0.4 & & 0.3 & & 0.7 &
& $0.82\pm0.17$ \cite{aoki2} & 2.598 & 2.598 & 2.597 & 2.622 \\

$^{15}_{\Xi s}$C & ${p\frac{1}{2}}^-\otimes{n\frac{1}{2}}^-\otimes{\Xi\frac{1}{2}}^+$
& 0 & 0 & 0 & 0 & 10.4 & (6.5) & 10.0 & (6.1) & 7.2 & (3.7)
& 9.4 (5.7) \cite{hu} & 2.582 & 2.539 & 2.545 & 2.597 \\

$^{15}_{\Xi p}$C & ${p\frac{1}{2}}^-\otimes{n\frac{3}{2}}^-\otimes{\Xi\frac{3}{2}}^-$
& 0 & -0.020 & -0.020 & -0.010 & 1.1 & & 1.1 & & 1.1 &
& 1.11$\pm$0.25 \cite{nakazawa1} & 2.582& 2.572 &2.576 & 2.592 \\

$^{17}_{\Xi s}$N & $0^+\otimes{\Xi\frac{1}{2}}^+$
& 0 & 0 & 0 & 0 & 11.4 & (7.1) & 11.2 & (6.9) & 8.3 & (4.3)
& 8.1 \cite{hu} & 2.676 & 2.634 & 2.639 & 2.686 \\

$^{17}_{\Xi p}$N & $0^+\otimes{\Xi\frac{3}{2}}^-$
& 0 & -0.005 & -0.005 &-0.004 & 2.4 & & 2.3 & & 2.2 &
& 2.1 \cite{hu} & 2.676 & 2.660 & 2.664 & 2.683 \\

$^{21}_{\Xi s}$F & $0^+\otimes{\Xi\frac{1}{2}}^+$
& 0.394 & 0.381 & 0.381 & 0.390 & 13.4 & (8.2) & 13.3 & (8.2) & 9.5 & (4.8)
& & 2.925 & 2.880 & 2.882 & 2.936 \\

$^{21}_{\Xi p}$F & ${p\frac{1}{2}}^-\otimes{\Xi\frac{3}{2}}^-$
& 0.394 & 0.377 & 0.377 & 0.381 & 3.5 & & 3.6 & & 3.1 &
& & 2.925 & 2.903 & 2.904 & 2.926 \\

$^{37}_{\Xi s}$Cl & $0^+\otimes{\Xi\frac{1}{2}}^+$
&-0.151 &-0.139 &-0.139 &-0.154 & 20.4 & (12.0) & 20.9 & (12.5) & 13.9 & (6.2)
& & 3.309 & 3.276 & 3.276 & 3.322 \\

$^{37}_{\Xi p}$Cl & ${p\frac{1}{2}}^-\otimes{\Xi\frac{3}{2}}^-$
&-0.151 &-0.158 &-0.158 &-0.154 & 12.4 & (4.8) & 12.6 & (5.0) & 9.3 & (2.3)
& & 3.309 & 3.293 & 3.293 & 3.318 \\

$^{41}_{\Xi s}$K & $0^+\otimes{\Xi\frac{1}{2}}^+$
& 0 & 0     & 0     & 0     & 21.3 & (12.3) & 21.9 & (12.8) & 15.0 & (6.7)
& 16.6 \cite{hu} & 3.398 & 3.368 & 3.368 & 3.407 \\

$^{41}_{\Xi p}$K & $0^+\otimes{\Xi\frac{3}{2}}^-$
%& 0 & 0 & 0 & 0 & 12.8 & (4.5) & 13.1 & (4.9) & 9.6 & (2.1)
& 0 &-0.003 &-0.003 &-0.001 & 13.3 & (5.0) & 13.6 & (5.3) & 10.2 & (2.5)
& 10.5 \cite{hu} & 3.398 & 3.378 & 3.378 & 3.404 \\

$^{91}_{\Xi s}$Y & $0^+\otimes{\Xi\frac{1}{2}}^+$
& 0 & 0     & 0     & 0     & 30.0 & (15.3) & 31.4 & (16.4) & 21.8 & (7.9)
& 23.0 \cite{hu} & 4.262 & 4.236 & 4.240 & 4.264 \\

$^{91}_{\Xi p}$Y & $0^+\otimes{\Xi\frac{3}{2}}^-$
%& 0 & 0 & 0 & 0 & 24.2 & (10.2) & 25.0 & (11.0) & 18.0 & (5.1)
& 0 & 0.003 & 0.003 & 0.003 & 24.2 & (10.2) & 25.1 & (11.1) & 18.4 & (5.5)
& 17.9 \cite{hu} & 4.262 & 4.245 & 4.245 & 4.266 \\

$^{209}_{\Xi s}$Tl & $0^+\otimes{\Xi\frac{1}{2}}^+$
& 0 & 0     & 0     & 0     & 40.1 & (16.5) & 41.6 & (17.9) & 31.8 & (8.8)
& 29.7 \cite{hu} & 5.543 & 5.526 & 5.526 & 5.546 \\

$^{209}_{\Xi p}$Tl & $0^+\otimes{\Xi\frac{3}{2}}^-$
%& 0 & 0 & 0 & 0 & 32.8 & (10.0) & 33.9 & (11.2) & 25.5 & (3.6)
& 0 & 0.001 & 0.001 & 0.001 & 36.0 & (13.5) & 37.3 & (14.7) & 28.6 & (7.1)
& 26.3 \cite{hu} & 5.543 & 5.529 & 5.529 & 5.544 \\

\end{tabular}
\end{ruledtabular}
\label{t:all}
%\mycom{SL3 radii are bigger !?}
\end{table*}%...................................................................

\begin{figure}[t]%..............................................................
\vspace{-2mm}
\centerline{\hspace{-0mm}\includegraphics[scale=0.52]{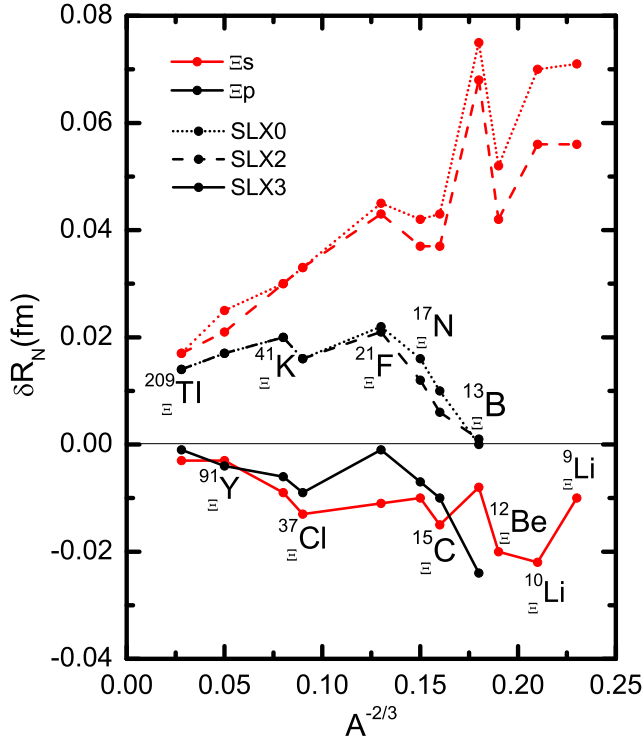}}
\vspace{-4mm}
\caption{
%Mass dependence of the
Core radius change $\delta R_N$,
Eq.~(\ref{e:dr}),
for several nuclei,
obtained with different $\nxi$ forces.
The subscripts $s$ and $p$ denote the $\xm$ orbit.}
\label{f:rad}
%\mycom{bigger legend order SLX3,SLX2,SLX0 \\}
\end{figure}%...................................................................

\begin{figure}[t]%..............................................................
\vspace{0mm}
\centerline{\hspace{0mm}\includegraphics[scale=0.48]{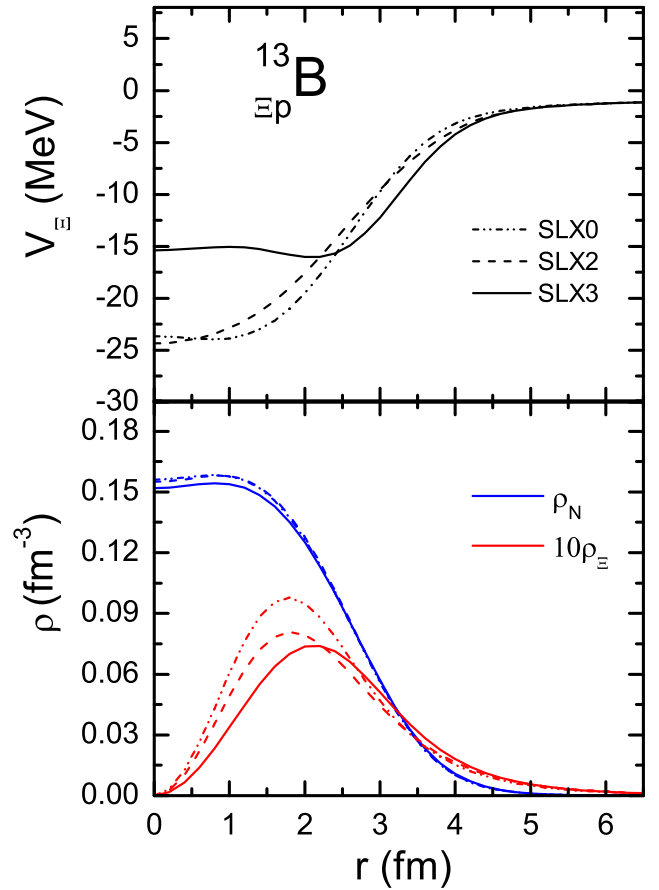}}
\vspace{-2mm}
\caption{
The mean field potential $V_\Xi$ (upper panel)
and the $N$ and $\Xi$ densities (lower panel)
in the hypernucleus \xibp\
calculated with different $\nxi$ forces.
}
\label{f:b}
\end{figure}%...................................................................

%-------------------------------------------------------------------------------
\section{Results}
\label{s:res}

We now study systematically a series of single \xim\ hypernuclei
in the ground state ($\Xi s$) and the first excited state ($\Xi p$).
Among them, the hypernuclei with spherical cores are
$^{15}_{\ \Xi}$C, $^{17}_{\ \Xi}$N, $^{41}_{\ \Xi}$K, $^{91}_{\ \Xi}$Y,
and $^{209}_{\ \ \ \Xi}$Tl;
and the ones with deformed cores are
$^9_\Xi$Li, $^{10}_{\ \Xi}$Li, $^{12}_{\ \Xi}$Be, $^{13}_{\ \Xi}$B,
$^{21}_{\ \Xi}$F, and $^{37}_{\ \Xi}$Cl.

\subsection{Radii}

In Table~\ref{t:all} we list
the calculated values of the deformation parameter $\beta_2$,
the $\xm$ separation energy $\bxi$,
and the nuclear core radius $R_N$ for the different \xim\ hypernuclei.
%It is found that the core deformation $\beta_2$ of a \xim\ hypernucleus
%in the ground state
%generally decreases compared to that of the core nucleus with all forces,
%the smallest deformation change effected by \sl3.
It is found that the core radius of a \xim\ hypernucleus
%in the ground state
generally decreases compared to that of the core nucleus
with the \sl0\ and \sl2\ forces,
but slightly increases with the \sl3.
This modification of the core size is larger
when the \xim\ hyperon occupies the
concentrated $1s$ state than the more diluted $1p$ state.

This is because
the magnitude of the core distortion
is determined by the competition
between the attractive linear term ($\sim a_0$)
and the repulsive nonlinear term ($\sim a_3$)
of the hyperonic SHF mean field Eq.~(\ref{e:vy}),
and thus depends on the $\nxi$ force,
see Fig.~1 of \cite{sun}.
The \sl3 force features strongest attraction at low density
in the peripheral region ($\ron\approx0.1$\fm3) of the core nucleus,
but much less at the higher central density
and thus tends to dilute the core,
whereas \sl0 or \sl2 exhibit the opposite behavior and contract the core.
For better illustration we plot in Fig.~\ref{f:rad} the radius change
\bea
% \delta R_N \equiv R_N[^{A+1}_{\quad\Xi}(Z-1)] - R_N[^AZ]
 \delta R_N \equiv R_N[^{A-1}(Z+1)] - R_N[^A_\Xi Z]
\label{e:dr}
\eea
for the different nuclei,
confirming the qualitative difference between the XNI forces
explained above,
namely, positive values of $\delta R_N$ with \sl0 and \sl2
and negative values with \sl3.
Moreover, the core radius changes are naturally largest for small nuclei.

\subsection{Removal energies}

These properties also influence the removal energies $\bxi$ listed in the table:
Apart from the weakly bound \xicp,
where all forces yield by construction the same
$\bxi=1.1\,$MeV,
the \sl3 force predicts generally smaller $\bxi$ values than \sl0 and \sl2
for larger and stronger bound hypernuclei,
where the relevant central high-density nuclear domain is more extended.
This feature allowed also the best fit of the experimental data for
the hypernucleus \xbes\ among the \sl* forces \cite{sun}.
The $\xm$ binding energies obtained by \sl3
are close to the results of the RMF model \cite{hu} as well.

However, for weakly bound hypernuclei like \xicp,
where a low-density nuclear environment prevails for the \xim,
the situation is opposite,
in particular
we notice that the \xim\ removal energy of the deformed nucleus
\xibp\ derived with \sl3 is 0.7 MeV,
larger than that of the other two forces.
This value is consistent with the experimental data of $0.82\pm0.17\;$MeV
in the E176 experiment \cite{nakazawa1,aoki2,emik}.
(The \xim\ hyperon occupies the $1p_{3/2}$ state.
In our model, the $[110]{1/2}^-$ and $[101]{3/2}^-$ orbits are the same
without hyperon spin-orbit coupling,
but the $[101]{1/2}^-$ orbit is separated due to the nuclear deformation).

In order to illustrate this hypernucleus in more detail,
we plot in Fig.~\ref{f:b}
the $\xm$ potential (upper panel)
and the $N,\Xi$ density distributions (lower panel).
With the \sl0,2,3 forces,
the depth of $V_\Xi(\rv=0)$ is about $23,24,15$ MeV.
However, the width of the \sl3 potential is larger,
thus providing more binding for the hyperon in the extended $p$ orbit:
It is seen in the lower panel that the major part of the hyperon
is located in the range $r\gtrsim2.5\,\text{fm}$,
where the \sl3 potential is deeper than those of \sl0,\sl2.

We notice another related interesting phenomenon regarding the
$\xm$ removal energy of deformed hypernuclei, namely
the results of $\bxi$ are different for deformed and undeformed calculations.
The quantity
\bea
 \Delta \bxi &\equiv& B_{\Xi}^\text{def.} - \bxi^\text{nondef.}
%\\\nonumber
% = \Delta E(^AZ) - \Delta E(^A_{\Xi}Z)
%\\\nonumber
% &=& [E(^AZ) - E(^A_KZ)]^\text{def.} - [E(^AZ) - E(^A_KZ)]^\text{nondef.}
\label{e:db}
\eea
is listed in Table~\ref{t:db} for various hypernuclei with sizeable deformations.
It is seen that the \xim\ hyperon in the $1s$ state is bound less in
deformed nuclei than the same undeformed ones,
which is caused by a reduction of the mean nucleon density in the former case.
However, for the $p$-state \xim\ hypernuclei
with oblate deformation
like \xibp\ and $^{\;37}_{\Xi p}$Cl,
the effect is opposite.
In these two hypernuclei,
the \xim\ hyperon occupying the extended $1p$ orbit is more bound
due to the core deformation,
which might actually increase the density in the peripheral part
where the hyperon resides.
In particular, the \xibp\ hypernucleus calculated by the 1D model
is unbound with
$\bxi=-0.27,-0.25,0.06\;$MeV in the cases of \sl0, \sl2, and \sl3, respectively,
while in the 2D approach it becomes bound with
$\bxi=0.43,0.27,0.68\;$MeV.
On the contrary,
in the prolately deformed hypernucleus \xifp,
the overlap of the $p$-state \xim\ hyperon with the prolate core
is weaker than in the undeformed case,
and therefore it is also less bound in the 2D model.

%The reason is that the deformation of the core nucleus $^{12}$C is oblate,
%just as the oblate $[110]{1/2}^-$ hyperon orbit in the hypernucleus,
%thus having the shrinking effect on the hyperon.
%%% this nucleus is not shrinking !!!

\begin{table}[t]%...............................................................
\caption{
The difference
between the 2D and 1D calculations
of the $\xm$ removal energy,
$\Delta\bxi = \bxi^{(2\text{D})} - \bxi^{(1\text{D})}$,
for several deformed nuclei
with different $\nxi$ forces.
}
\renewcommand{\arraystretch}{1.2}
\begin{ruledtabular}
\begin{tabular}{l|d|ddd}
& \beta_2 &\multicolumn{3}{c}{$\Delta\bxi$ (MeV)} \\
& \text{no }\Xi & \text{\sl0} & \text{\sl2} & \text{\sl3} \\
\hline
$^{\ 9}_{\Xi s}$Li  &  0.62 & -0.70 & -0.63 & -0.08  \\
$^{10}_{\Xi s}$Li   &  0.31 & -0.22 & -0.21 & -0.01 \\
$^{12}_{\Xi s}$Be   &  0.13 & -0.04 & -0.05 &  0.00 \\
$^{13}_{\Xi s}$B    & -0.30 & -0.51 & -0.49 &  0.00\\
$^{13}_{\Xi p}$B    & -0.30 &  0.70 &  0.52 &  0.62 \\
$^{21}_{\Xi s}$F    &  0.39 & -0.19 & -0.27 & -0.16 \\
$^{21}_{\Xi p}$F    &  0.39 & -1.08 & -0.98 & -0.68 \\
$^{37}_{\Xi s}$Cl   & -0.15 & -0.25 & -0.24 & -0.06 \\
$^{37}_{\Xi p}$Cl   & -0.15 &  0.41 &  0.42 &  0.34 \\
\end{tabular}
\end{ruledtabular}
\label{t:db}
\end{table}%....................................................................

\begin{figure}[t]%..............................................................
\vspace{-2mm}
\centerline{\hspace{-4mm}\includegraphics[scale=0.38]{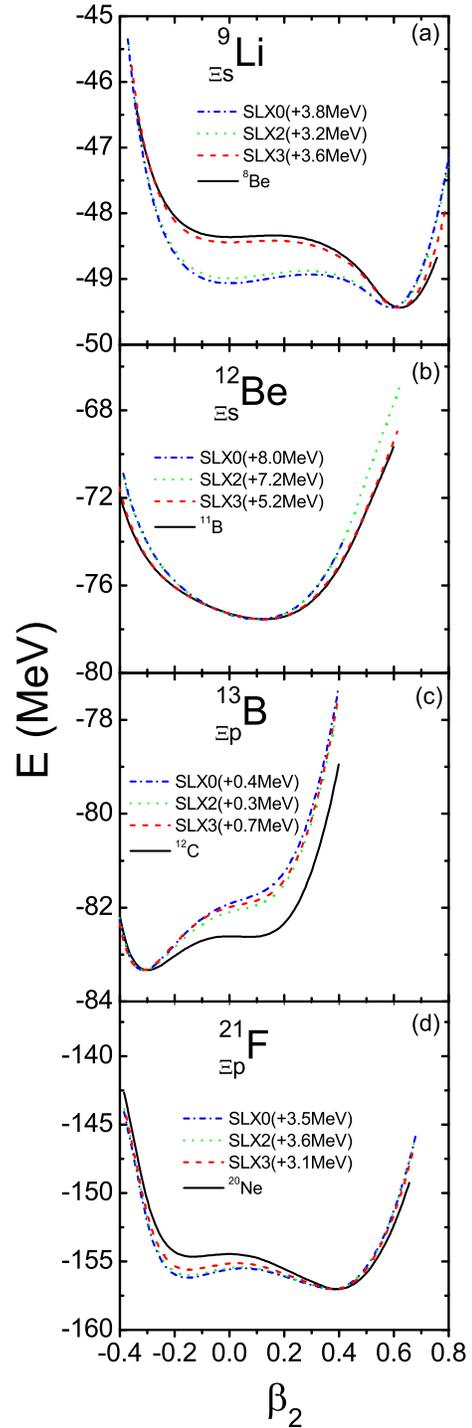}}
\vspace{-5mm}
\caption{
The binding energy surfaces
obtained with different $\nxi$ forces
for
(a) $^8$Be and $^{\ 9}_{\Xi s}$Li;
(b) $^{11}$B and \xbes;
(c) $^{12}$C and \xibp;
(d) $^{20}$Ne and \xifp.
Note the different energy scales.
}
\label{f:bes}
\end{figure}%...................................................................

\begin{figure}[t]%..............................................................
\vspace{-1mm}
\centerline{\hspace{-2mm}\includegraphics[scale=0.55]{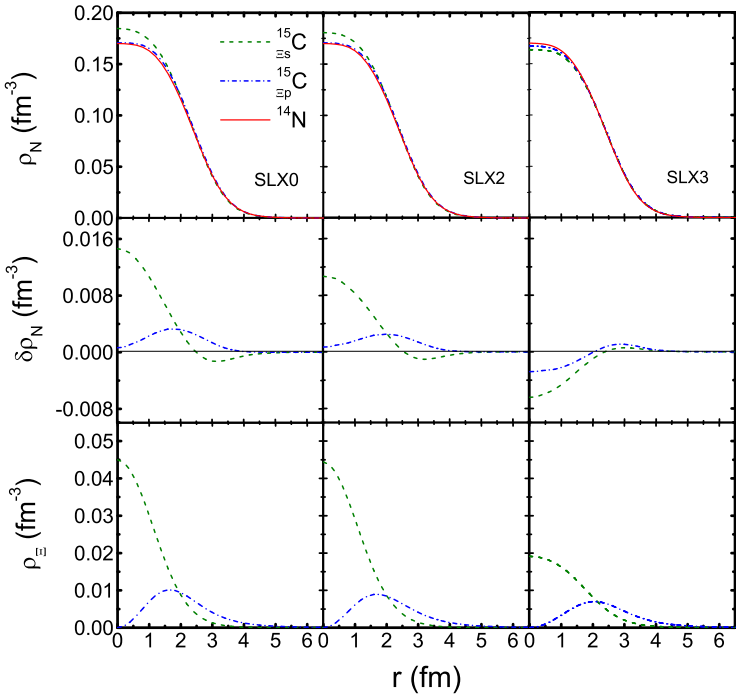}}
\vspace{-2mm}
\caption{
Nucleon and hyperon density distributions of \xics, \xicp,
and the core nucleus $^{14}$N,
obtained with different $\nxi$ forces.
$\delta\ron$ is the change of the nucleon density by the added hyperon.
The subscripts $s$ and $p$ denote the $\xm$ orbit.
}
\label{f:rho}
\end{figure}%...................................................................

\begin{figure}[t]%..............................................................
\vspace{-0mm}
\centerline{\hspace{0mm}\includegraphics[scale=0.49]{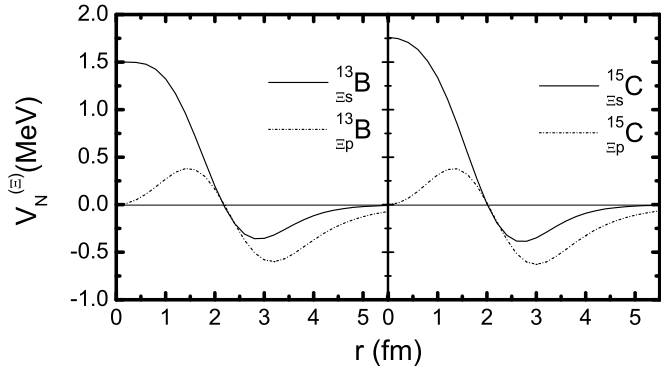}}
\vspace{-3mm}
\caption{
The hyperonic contribution to the nucleon mean field $V_N^{(\Xi)}$,
Eq.~(\ref{e:vnx}),
for the $s$- or $p$-state hypernuclei \xib\ and \xic\
with the \sl3 force.
}
\label{f:pot}
\end{figure}%...................................................................

\begin{figure}[t]%..............................................................
\vspace{-1mm}
\centerline{\hspace{0mm}\includegraphics[scale=0.49]{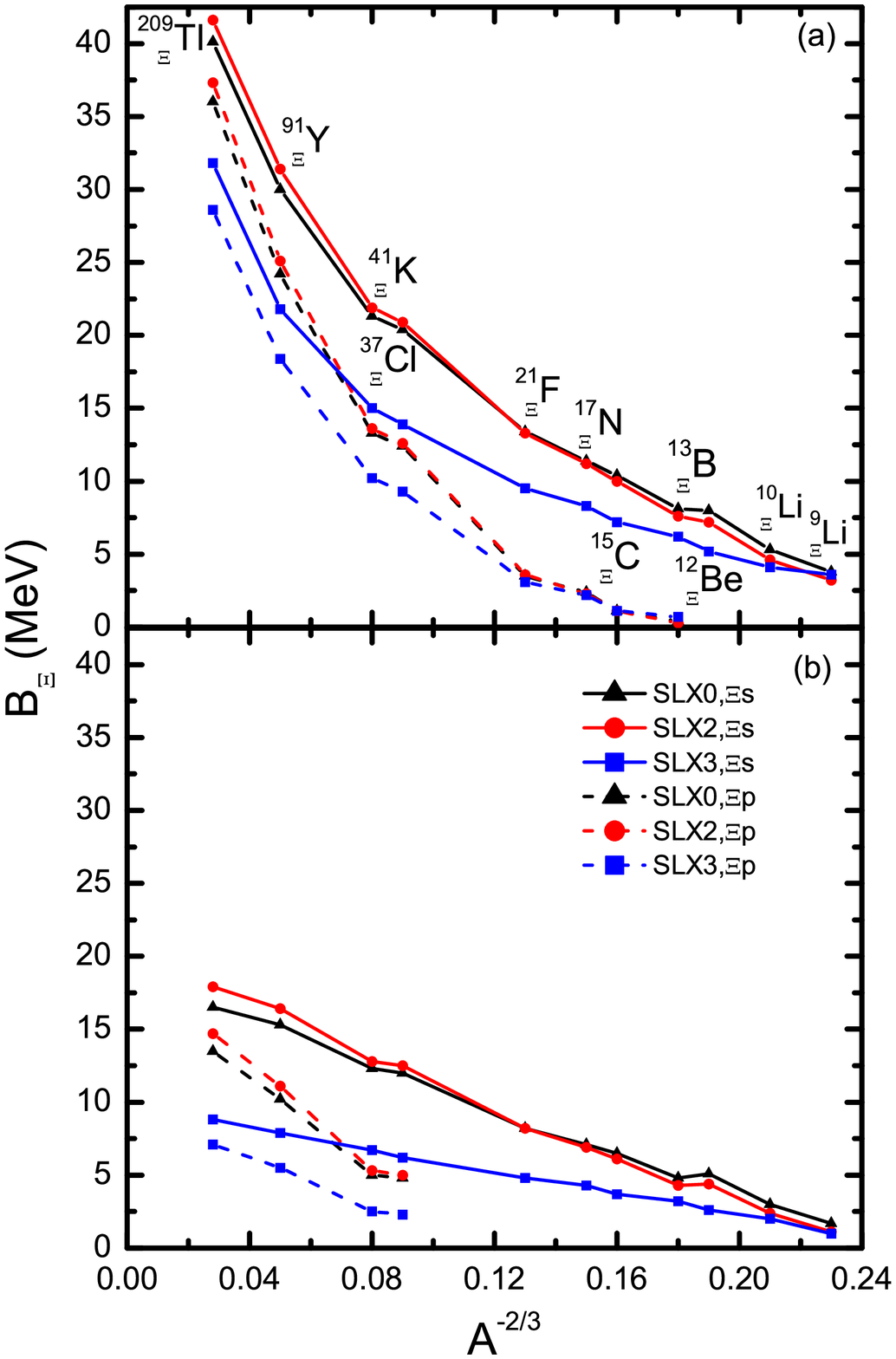}}
\vspace{-4mm}
\caption{
Hyperon binding energies $B_\Xi$ as a function of mass number $A^{-2/3}$
with different $\nxi$ forces and
with (a) or without (b) $\nxi$ Coulomb potential.
}
\label{f:ba}
\end{figure}%...................................................................

\subsection{Deformation}

To study in more detail the effect of a single \xim\ hyperon
on the nuclear core deformation,
we plot in Fig.~\ref{f:bes}
the binding energy surfaces (BESs) of the core nuclei
$^8$Be, $^{11}$B, $^{12}$C, $^{20}$Ne,
and their \xim\ hypernuclei
$^{\ 9}_{\Xi s}$Li, $^{12}_{\Xi s}$Be, \xibp, \xifp.
In order to better compare the curves,
the BESs of the \xim\ hypernuclei are shifted by a constant amount
so as to obtain the same minimum value.
The corresponding values of $\beta_2$ at minimum are listed in
Tables~\ref{t:all} and \ref{t:db}.
We observe that
in general the addition of a hyperon favors oblate deformations,
which in Fig.~\ref{f:bes}(a,c,d)
become more advantageous relative to the prolate ones.
Thus an eventual secondary minimum becomes more pronounced (a,d)
or disappears entirely (c).

These effects can be understood by analyzing the overlap of the
density distribution of the added \xim\ hyperon
with the nucleon density distributions of configurations of various shapes,
which determines the energy gain by the added hyperon.
For example, returning to the case of \xibp\ in panel (c),
one observes that due to the overlap of the $\Xi^-_p$ density distribution
and the deformed nucleon distribution
the total energy is lowered for oblate configurations
but increased for prolate ones.
Consequently the existing primary oblate minimum becomes even more pronounced.
The same effect is exhibited for
$^{\ 9}_{\Xi s}$Li~(a) and
\xifp~(d),
where the secondary, oblate, minima become deeper when adding the hyperon.
In particular for the light nucleus $^{\ 9}_{\Xi s}$Li~(a),
we notice that the BESs by the three forces show visible differences,
particularly for \sl3,
whose curve maintains the most pronounced minimum
compared to the other two curves.

\subsection{Densities}

More information on the
nucleon and hyperon density distributions is given in
Fig.~\ref{f:rho},
which shows those of the hypernucleus \xic\ obtained with the different XNIs.
Consistent with the analysis of the nuclear radii in Table~\ref{t:all},
one observes that the central nucleon density $\ron(\rv=0)$
in hypernuclei calculated by \sl0 and \sl2 is larger than that in the core nuclei
(core shrinking),
while the opposite occurs with the force \sl3
(core expansion).
For the $p$-state hypernucleus \xicp,
the effect is reduced due to the vanishing of the hyperon density in the core,
and consequently in Fig.~\ref{f:rho} with the \sl3 force,
the central nucleon density of the $p$-state hypernuclei
is larger than that of the $s$-state hypernuclei,
while all forces make more nucleons reside in the peripheral region
%($\ron \approx 0.075-0.15$\fm3)
($r \gtrsim 2\,\text{fm}$)
of $p$-state hypernuclei.

These modifications of the nucleonic density
can be understood by the hyperonic part of the nucleon mean field
$V^{(\Xi)}_N \sim (a_0+2a_3\ron)\roy$,
which due to the competition between
the attractive $a_0$ and repulsive $a_3$ components
for the \sl3 force is repulsive at the center
and attractive in the peripheral region
($r \gtrsim 2\,\text{fm}$)
of the \xic\ nucleus,
as shown in Fig.~\ref{f:pot}.
With the current parameter set,
the transition between repulsion and attraction occurs at
$\ron = -a_0/2a_3 = 0.114$\fm3.

Consistent effects on the hyperon density are observed
in the lower panel of Fig.~\ref{f:rho},
namely the hyperon density is more spread out and smaller in the center
with the force \sl3\ than with the other two forces.
On a reduced scale,
these effects also take place in the heavier hypernuclei examined.
%$^{41}_{\ \Xi}$K and $^{209}_{\ \ \;\Xi}$Tl.

\subsection{\xim\ binding}

Finally,
Figure~\ref{f:ba} summarizes the mass dependence of the hyperon removal energies
obtained with three XNIs \sl0, \sl2 and \sl3,
including or not the $\nxi$ Coulomb interaction,
which becomes increasingly important for the heavier nuclei
due to its long-range nature.
For heavy nuclei the results with the full interaction
approach about 30 MeV with the \sl3 force,
which is very similar to the case of $\Lambda$ hypernuclei \cite{hypsky}.
Due to the missing repulsive contribution at normal density,
the \sl0 or \sl2 force yield about 10 MeV more binding for heavy nuclei.

%-------------------------------------------------------------------------------
\section{Conclusions}
\label{s:end}

We systematically studied
hyperon removal energies, nuclear core radii, core deformations,
and density distributions
of \xim\ hypernuclei using a 2D SHF model
with three different $\nxi$ Skyrme forces
adjusted to reproduce experimental data for \xicp.

The \sl3 $\nxi$ force is the only one to yield simultaneous
agreement with the \xbes\ data.
It incorporates an important density-dependent
repulsive component
and thus yields maximum attraction at subnuclear densities.
Consequently
the induced core deformation is a dilatation
and the $\xm$ binding energies of heavy nuclei
obtained with \sl3 are smaller than those with the other forces,
and more similar to those of RMF or cluster model calculations.

The hyperon removal energy of \xibp\ with the \sl3 force is 0.7 MeV,
compatible with a possible value of $0.82\pm0.17$ MeV
interpreted from the KEK E176 experiment.
We thus predict that this system is a possibly bound
$p$-state \xim\ hypernucleus.
But still more accurate experiments are needed to confirm the reliability
of the observed event.
Also, the prediction of this weakly bound state is delicate,
as it depends on the nuclear deformation:
This hypernucleus is unbound in the 1D model,
but bound in the 2D model
due to its more extended geometry favoring lower and more attractive
nuclear densities.

This feature was studied in detail analyzing the
nuclear core radii and density distributions of several \xim\ hypernuclei.
Due to their different density dependence,
the three $\nxi$ Skyrme forces impose different variations
on the nucleon density compared to that of the core nuclei:
With \sl0 or \sl2, the central nucleon density increases,
while it decreases with \sl3
due to a repulsive effect on the nucleons in the central region of nuclei.
This leads to a contraction/dilatation of the core, respectively.

These differences also influence the BESs of the various nuclei,
and we have pointed out the delicate interplay between
\xim\ $s$ or $p$ state occupation
and the prolate or oblate nuclear environment,
which modifies the minima of the BES.

In this context we remark, however,
that the mean-field approximation employed here
might be inadequate in particular for weak BES minima,
due to the neglect of configuration mixing.
A beyond-mean-field treatment \cite{cui,mei}
might be required for a more realistic modelling in the future.

Finally, we studied the mass dependence of the \xim\ binding energies
in a broad region of the mass table
and predict removal energies of about 30 MeV for the heaviest hypernuclei
with the \sl3 force.

We hope that our calculations and predictions can help to identify
the bound \xim\ hypernuclei.
Currently, there are several projects for measuring \xim\ hypernuclei
and \xim\ atoms by upgraded technologies at J-PARC and other facilities.
In the future,
more reliable and accurate experimental data on \xim\ hypernuclei
will become available and
provide more constraints on the $\nxi$ interaction.
This will allow to determine more precisely the parameters of the $\nxi$
Skyrme force
and permit more accurate predictions.

\phantom{.}\vspace{0mm}
%-------------------------------------------------------------------------------
\section*{Acknowledgments}

We thank Ji-Wei Cui for suggestive discussions.
This work was supported by the National Science Foundation of China
under contract No.~11775081,
and the Natural Science Foundation of Shanghai under contract No.~17ZR1408900.

%-------------------------------------------------------------------------------

\end{document}